\documentclass[sigchi]{acmart} 

\AtBeginDocument{%
  }

\setcopyright{acmlicensed}
\copyrightyear{2025}
\acmYear{2025}
\acmDOI{XXXXXXX.XXXXXXX}
\acmConference[CHI '25]{CHI Conference on Human Factors in Computing Systems}{Apr 26 -- May 1, 2025}{Yokohama, Japan}

\usepackage{subcaption} 
\usepackage{url}
\usepackage{xcolor}
\usepackage{booktabs}
\usepackage{tabularx}
\usepackage{verbatim}
\makeatletter
\g@addto@macro{\UrlBreaks}{\UrlOrds}
\makeatother
\usepackage{multirow} 
\usepackage{color} 
\usepackage{enumitem} 

\usepackage{longtable}
\usepackage{xspace}
\usepackage{tcolorbox}
\usepackage{hyperref}
\usepackage{todonotes}

\newtcbox{\highlight}[1][magenta]{on line, arc=0pt,colback=#1!10!white,colframe=#1!50!black, before upper={\rule[-3pt]{0pt}{10pt}},boxrule=1pt, boxsep=0pt,left=6pt,right=5pt,top=4pt,bottom=3pt}

\newcommand{\notextrefrq}[1]{\hyperref[rq:#1]{\highlight{\textbf{\texttt{\textcolor{gray}{#1}}}}}}

\usepackage{todonotes}

\usepackage{duckuments}

\begin{document}

\title{Community Empowerment through Location-Based AR: The Thámien Ohlone AR Tour}

\author{Kai Lukoff}
\email{klukoff@scu.edu}
\orcid{0000-0001-5069-6817}
\affiliation{%
  \institution{Santa Clara University}
  \city{Santa Clara}
  \state{CA}
  \country{USA}
}

 \author{Xinqi Zhang}
\email{xzhang22@scu.edu}
\orcid{0009-0004-7493-9800}
\affiliation{%
  \institution{Santa Clara University}
  \city{Santa Clara}
  \state{CA}
  \country{USA}
}

\begin{abstract}
Community empowerment is the process of enabling communities to increase control over their narratives, resources, and futures. In HCI and design, this social challenge centers on helping marginalized groups gain agency through technology and design interventions. For Indigenous communities in particular, empowerment means not only representation but sovereignty in how their stories are told and by whom. Location-based augmented reality (AR) offers a novel opportunity to address this challenge. By overlaying digital content onto physical places, AR can spatially anchor community narratives in the real world, allowing communities to re-tell the story of a place on their own terms. Such site-specific AR experiences have already been used to reveal hidden histories, re-imagine colonial monuments, and celebrate minority cultures. The affordances of XR – particularly AR’s spatial interaction and immersive storytelling – make it a promising tool for cultural continuity and community activism. In this position paper, we focus on how these XR affordances can empower communities, using the Thámien Ohlone AR Tour as a case study. We outline why traditional digital interventions fall short of true empowerment, how AR’s immersive qualities uniquely support Indigenous self-determination, insights from co-designing the Ohlone AR Tour, and future directions to scale such efforts responsibly.
\end{abstract}

\maketitle

\section{Current Limitations of Digital Interventions for Empowerment}
Existing digital solutions often inadequately address community empowerment, especially for Indigenous groups. Traditional heritage apps, websites, or museum exhibits typically present content about communities but are created without meaningful community leadership. This top-down approach means that marginalized communities remain subjects rather than co-creators, limiting any empowerment benefits \cite{Costanza-Chock2020-bl}. Simply “including” community input is not enough if power over design decisions and narratives stays with external experts. As a result, many digital interventions fail to build local capacity or shift representation in a lasting way. 

In the context of AR and XR, additional barriers emerge. Deploying location-based AR experiences usually requires technical resources, stable platforms, or permission from authorities – factors that communities often do not control \cite{Silva2024-jr}. For example, activists wishing to overlay Indigenous histories onto a city landscape might need to use proprietary AR platforms or seek approval from landowners, which can constrain grassroots action. Moreover, current AR heritage projects can suffer from scalability and sustainability issues: each experience tends to be custom-built and maintenance-intensive, making it hard for small communities to adopt long-term. Accessibility is another limitation – advanced XR often assumes users have high-end devices and digital literacy, which might exclude some community members. In short, today’s digital interventions rarely hand over the “keys” of design and storytelling to the communities themselves. This gap points to the need for new approaches where technology development is steered by those it intends to empower, aligning with calls for design justice and decolonizing design methods. Without such shifts, digital projects risk reinforcing the very power imbalances they aim to fix, offering content to communities but not genuine capacity or agency.

\section{XR Affordances for Community Empowerment}
XR technologies – especially augmented reality – provide unique affordances that can support community empowerment in ways traditional media cannot. Two key affordances are spatial interaction and immersion, which together allow for deeply contextual and engaging experiences:
\begin{itemize}
        \item \textbf{Place-Based Storytelling}: AR’s ability to tie digital content to specific physical locations enables narratives to be experienced in situ. For Indigenous communities, whose stories and knowledge are often profoundly linked to the land, this spatial anchoring is critical. AR can reinforce land-based epistemologies, meaning that teachings and histories are embedded in the places where they occurred. For example, standing on an ancestral village site and viewing a digitally reconstructed scene or hearing an elder’s story in that exact spot creates a powerful connection between the narrative and the landscape. This affordance helps ensure Indigenous histories remain linked to their rightful geographic context, counteracting the “placelessness” that sometimes comes with digital media. By reclaiming public spaces with their own stories, communities exercise a form of spatial empowerment, asserting that “We are still here” \cite{Jauregui2024-ai}.
        \item \textbf{Immersion and Affective Engagement}: The immersive nature of XR, even via a phone-based AR view, can evoke a strong sense of presence and emotional resonance. Unlike reading text or viewing static images, AR immerses users in an embodied narrative experience – one sees and hears Indigenous perspectives overlaid on the immediate surroundings. For the community, creating immersive AR content can also be empowering internally. It allows Indigenous creators to present their culture dynamically, using 3D visuals, audio in their own languages, and interactive elements that bring pride and a sense of modern presence. Immersion turns passive audiences into active participants who must move through real-world spaces, uncover stories, and reflect on their relationship to that space. Such embodied interaction aligns with Indigenous ways of knowing that emphasize learning through experience and reflection in place.
\end{itemize}

Together, these affordances of spatial interaction help ensure the technology serves the community’s connection to their land, and immersion ensures the message resonates deeply with users.

\subsection{Case Study: The Thámien Ohlone AR Tour}
To illustrate how location-based AR can empower a community, we present the Thámien Ohlone AR Tour – a project co-designed by the Muwekma Ohlone Tribe and Santa Clara University (SCU). This project is rooted in the Santa Clara Valley (California), on lands traditionally known as Thámien, which were inhabited by Ohlone people for millennia before Spanish colonization. Today, the site hosts Mission Santa Clara and SCU’s campus, places that often obscure the Indigenous history of the land. The AR tour was conceived to highlight the often-overlooked Ohlone presence on campus and to educate both the public and the SCU community about Ohlone history, contemporary culture, and ongoing presence. Importantly, from its inception, the project was driven by the Indigenous community’s vision of how their story should be told. Co-Design Process: The Thámien Ohlone AR Tour was not built by outsiders and “handed” to the tribe; it was co-designed with the community at every step. Initial concept workshops began in 2018, initiated by Ohlone tribal members, and formal co-design sessions took shape through 2020 and 2021. The development team included Muwekma Ohlone elders, youth, and university researchers working together as equal stakeholders. This collaborative process followed an “I Do, We Do, You Do” pedagogical model, emphasizing reciprocal learning. In practical terms, the co-design involved:
\begin{itemize}
        \item \textbf{Storytelling Workshops}: Community elders and historians led sessions to decide which stories, sites, and themes should be featured in the AR tour. Rather than a researcher dictating content, Ohlone participants determined what narratives were most important to convey – for example, creation stories, experiences of missionization, or contemporary cultural practices. These workshops ensured the content was grounded in Indigenous perspectives and priorities, reflecting the principle of rhetorical sovereignty \cite{Lyons2000-kl}.
        \item \textbf{AR Skills Training}: Ohlone youth and other community members received hands-on training in AR development technologies. Through biweekly workshops, university mentors taught skills like 3D modeling, Unity game engine use, and usage of AR platforms (e.g., Niantic’s Lightship visual positioning system). Community members learned to create digital assets — for instance, designing a 3D model of a traditional Ohlone village or programming an AR interaction. This skill-building was a core empowerment goal: participants were gaining tech literacy that could be applied beyond this project.
        \item \textbf{Iterative Prototyping with Feedback}: Rather than a one-off development, the AR content went through multiple iterations, with frequent feedback from the Ohlone co-designers and broader tribe. Early prototypes of each “tour stop” were tested by community members to ensure cultural accuracy, appropriateness, and a meaningful user experience. This iterative loop meant the final product truly reflected community input and could be adjusted if something didn’t feel right to Ohlone partners. 
\end{itemize}
By May 2023, the team publicly launched the first version of the Thámien Ohlone AR Tour with four initial locations on campus. Each AR “stop” in the tour corresponds to a significant site and story co-chosen by the community. For example, one stop is Thámien Village, where visitors using the AR app can view a reconstructed digital model of the precolonial Ohlone village that once stood on that ground. Another stop, Names in the Sky, projects the names of Ohlone ancestors (who are buried in an unmarked mission cemetery on campus) into the sky as a moving tribute. There is also an AR representation of the annual Powwow held at SCU and an overlay showing Ohlone basket-weaving technology in a modern context. All content is narrated by Ohlone voices, so users hear directly from tribal elders and youth rather than outside experts. By anchoring these stories to physical campus landmarks, the tour creates a visible Indigenous presence in a space that often marginalizes it. As one tribal member noted, the project communicates “This is Ohlone land” to all who experience it \cite{Jauregui2024-ai}. 

The co-design approach turned the AR tour into a vehicle for empowerment even before any end-user ever opened the app. Over the course of the project, Ohlone youth became content creators and technologists, developing competencies in 3D modeling, spatial computing, user experience design, and AR development. These skills not only contributed to the AR tour but also enhanced the youths’ capacity for future projects (be it cultural preservation or career opportunities in tech). Just as importantly, the project facilitated intergenerational knowledge exchange: elders passed down oral histories and cultural knowledge to the younger members during the design of the tour content. In this way, designing the AR experience doubled as a form of cultural education and youth engagement for the community. 

Furthermore, the AR tour has strengthened community bonds and public presence through events and ongoing collaboration. The initial launch and subsequent updates were celebrated at community gatherings: for instance, new AR tour stops were unveiled during Indigenous People’s Day 2023 and at the university’s spring Powwow event. These public launch events combined traditional ceremonies (dance, music, prayer) with live demonstrations of the AR app, turning the typically solitary act of using a phone app into a shared cultural celebration. The technology thus became a focal point for community events, prompting dialogues between Ohlone leaders, students, faculty, and local residents about Ohlone history and sovereignty. By integrating digital storytelling with real-world community action, the project exemplified what Atalay calls using technology as an “avenue for Indigenous cultural practice, rather than a site of knowledge extraction” \cite{Atalay2006-nz}. 

\section{Future Directions and Challenges}
Our work on the Thámien Ohlone AR Tour points toward promising future directions for XR in community empowerment, but it also highlights important technological needs, potential risks, and research hurdles for scaling up such efforts. 

\textbf{Scaling Technological Infrastructure}: For location-based AR to empower communities broadly, improvements in AR infrastructure are necessary. Current projects rely on platforms like Niantic Lightship or 8th Wall for mapping and anchoring content, which may not be accessible or affordable to all groups. There is a need for more open and robust AR mapping systems that communities can use with greater autonomy. For example, research into persistent AR anchors and slam (Simultaneous Localization and Mapping) techniques in varying environments will help – communities should be able to create stable AR content in outdoor settings like parks or remote areas without specialized equipment. Additionally, hardware advances (e.g. widespread AR glasses or improved mobile cameras) could make these experiences more seamless and immersive. However, those must be coupled with affordability; otherwise, only well-funded groups can leverage them. In that vein, one intended outcome of this project is to create a toolkit for location-based AR for social impact: by packaging templates and guides for AR storytelling, they lower the technical barrier and make replication easier. In the future, we envision community members using such toolkits to create their own AR tours with minimal expert help – akin to how easy blogging platforms enabled grassroots journalism. 

\textbf{Balancing Specificity and Replicability}: Each community has unique histories, cultural protocols, and needs, which means empowerment projects must be tailored – a one-size-fits-all approach will not work. Kari Noe and colleagues note that generalized “equitable design” frameworks often fail unless adapted to the specific Indigenous community’s context \cite{Noe2024-wq}. This suggests future XR projects should prioritize extensive preliminary engagement to understand local values and knowledge systems before design even begins. However, this emphasis on specificity can conflict with the desire to scale up solutions for broader impact. A research hurdle is finding what aspects of projects like the Ohlone AR Tour are generalizable (e.g., participatory methods, certain technical tools) and what must be re-invented with each new community. The AR Toolkit mentioned earlier attempts to strike this balance by sharing core techniques but encouraging flexible application. Further work in HCI could develop libraries of community empowerment patterns – design patterns or guidelines that future teams can draw on, while case studies continue to expand our understanding of diverse contexts. 

In summary, the road ahead for scaling location-based AR in service of community empowerment involves intertwining technical innovation with deep community collaboration. We must advance the technology to be more accessible, persistent, and secure for communities, and advance our participatory design methods to be more inclusive of indigenous epistemologies and protective of community rights. By anticipating risks – from cultural appropriation to platform dependency – we can proactively design for an ethical future of XR. The promise, as shown by the Thámien Ohlone AR Tour, is that when done right, XR can move beyond novel interfaces to become a platform for social justice and self-determination. The lesson for the HCI community is that the affordances of XR are powerful tools, but only if wielded in partnership with the people whose stories we seek to amplify.

\begin{acks}
We thank our Ohlone collaborators. This position paper was presented at the CHI 2025 workshop (arXiv:2504.07475).
\end{acks}

\bibliographystyle{ACM-Reference-Format}

\begin{thebibliography}{6}


\ifx \showCODEN    \undefined \def \showCODEN     #1{\unskip}     \fi
\ifx \showISBNx    \undefined \def \showISBNx     #1{\unskip}     \fi
\ifx \showISBNxiii \undefined \def \showISBNxiii  #1{\unskip}     \fi
\ifx \showISSN     \undefined \def \showISSN      #1{\unskip}     \fi
\ifx \showLCCN     \undefined \def \showLCCN      #1{\unskip}     \fi
\ifx \shownote     \undefined \def \shownote      #1{#1}          \fi
\ifx \showarticletitle \undefined \def \showarticletitle #1{#1}   \fi
\ifx \showURL      \undefined \def \showURL       {\relax}        \fi
\providecommand\bibfield[2]{#2}
\providecommand\bibinfo[2]{#2}
\providecommand\natexlab[1]{#1}
\providecommand\showeprint[2][]{arXiv:#2}

\bibitem[Atalay(2006)]%
        {Atalay2006-nz}
\bibfield{author}{\bibinfo{person}{Sonya Atalay}.} \bibinfo{year}{2006}\natexlab{}.
\newblock \showarticletitle{Indigenous archaeology as decolonizing practice}.
\newblock \bibinfo{journal}{\emph{American Indian quarterly}} \bibinfo{volume}{30}, \bibinfo{number}{3} (\bibinfo{year}{2006}), \bibinfo{pages}{280--310}.
\newblock
\showISSN{0095-182X,1534-1828}
\href{https://doi.org/10.1353/aiq.2006.0015}{doi:\nolinkurl{10.1353/aiq.2006.0015}}


\bibitem[Costanza-Chock(2020)]%
        {Costanza-Chock2020-bl}
\bibfield{author}{\bibinfo{person}{Sasha Costanza-Chock}.} \bibinfo{year}{2020}\natexlab{}.
\newblock \bibinfo{booktitle}{\emph{Design Justice: Community-Led Practices to Build the Worlds We Need}}.
\newblock \bibinfo{publisher}{MIT Press}.
\newblock
\showISBNx{9780262043458}
\urldef\tempurl%
\url{https://play.google.com/store/books/details?id=h4LPDwAAQBAJ}
\showURL{%
\tempurl}


\bibitem[Jauregui et~al\mbox{.}(2024)]%
        {Jauregui2024-ai}
\bibfield{author}{\bibinfo{person}{Cinthya Jauregui}, \bibinfo{person}{Tiffany~T Nguyen}, \bibinfo{person}{Sarah~Hazel Sallee}, \bibinfo{person}{Mohan~Raj Chandrasekar}, \bibinfo{person}{Liam A'Hearn}, \bibinfo{person}{Dominic~Jonathan Woetzel}, \bibinfo{person}{Pinak Paliwal}, \bibinfo{person}{Shea MacDonald}, \bibinfo{person}{Madison Nguyen}, \bibinfo{person}{Lee~M Panich}, \bibinfo{person}{Danielle~M Heitmuller}, \bibinfo{person}{Amy Lueck}, {and} \bibinfo{person}{Kai Lukoff}.} \bibinfo{year}{2024}\natexlab{}.
\newblock \showarticletitle{We Are Still Here: The Thámien Ohlone Augmented Reality Tour}. In \bibinfo{booktitle}{\emph{Extended Abstracts of the 2024 {CHI} Conference on Human Factors in Computing Systems}}. \bibinfo{publisher}{ACM}, \bibinfo{address}{New York, NY, USA}.
\newblock
\href{https://doi.org/10.1145/3613905.3649127}{doi:\nolinkurl{10.1145/3613905.3649127}}


\bibitem[Lyons(2000)]%
        {Lyons2000-kl}
\bibfield{author}{\bibinfo{person}{Scott~Richard Lyons}.} \bibinfo{year}{2000}\natexlab{}.
\newblock \showarticletitle{Rhetorical Sovereignty: What Do American Indians Want from Writing?}
\newblock \bibinfo{journal}{\emph{College Composition and Communication}} \bibinfo{volume}{51}, \bibinfo{number}{3} (\bibinfo{year}{2000}), \bibinfo{pages}{447--468}.
\newblock
\showISSN{0010-096X}
\href{https://doi.org/10.2307/358744}{doi:\nolinkurl{10.2307/358744}}


\bibitem[Noe and Kirshenbaum(2024)]%
        {Noe2024-wq}
\bibfield{author}{\bibinfo{person}{Kari Noe} {and} \bibinfo{person}{Nurit Kirshenbaum}.} \bibinfo{year}{2024}\natexlab{}.
\newblock \showarticletitle{Where Generalized Equitable Design Practice Meet Specific Indigenous Communities}. In \bibinfo{booktitle}{\emph{Proceedings of the CHI Conference on Human Factors in Computing Systems}} \emph{(\bibinfo{series}{CHI '24}, \bibinfo{number}{Article 616})}. \bibinfo{publisher}{Association for Computing Machinery}, \bibinfo{address}{New York, NY, USA}, \bibinfo{pages}{1--8}.
\newblock
\showISBNx{9798400703300}
\href{https://doi.org/10.1145/3613904.3642931}{doi:\nolinkurl{10.1145/3613904.3642931}}


\bibitem[Silva et~al\mbox{.}(2024)]%
        {Silva2024-jr}
\bibfield{author}{\bibinfo{person}{Rafael M~L Silva}, \bibinfo{person}{Ana~María Cárdenas~Gasca}, \bibinfo{person}{Joshua~A Fisher}, \bibinfo{person}{Erica Principe~Cruz}, \bibinfo{person}{Cinthya Jauregui}, \bibinfo{person}{Amy Lueck}, \bibinfo{person}{Fannie Liu}, \bibinfo{person}{Andrés Monroy-Hernández}, {and} \bibinfo{person}{Kai Lukoff}.} \bibinfo{year}{2024}\natexlab{}.
\newblock \showarticletitle{With or Without Permission: Site-Specific Augmented Reality for Social Justice}. In \bibinfo{booktitle}{\emph{Extended Abstracts of the 2024 {CHI} Conference on Human Factors in Computing Systems}}. \bibinfo{publisher}{ACM}, \bibinfo{address}{New York, NY, USA}.
\newblock
\href{https://doi.org/10.1145/3613905.3636283}{doi:\nolinkurl{10.1145/3613905.3636283}}


\end{thebibliography}

\end{document}